# AI-Driven Media & Synthetic Knowledge: Rethinking Society in Generative Futures

EXPLORING GENERATIVE AI FUTURES THROUGH AN EXPERIMENTAL PHD SEMINAR

Katalin FEHER
feher.katalin@uni-nke.hu
Ludovika University of Public Service

**Abstract**
Generative AI is not just a technological leap—it is a societal stress test, reshaping trust, identity, equity, and authorship. This exploratory PhD seminar examined emerging academic trends in AI-driven synthetic media and worlds, emphasizing ethical risks and societal implications. In Part One, students explored core concepts such as generative AI, fake media, and synthetic knowledge production. In Part Two, they critically engaged with these challenges, producing actionable insights. The two-part format enabled deep reflection on power, responsibility, and education in AI-augmented communication. Outcomes offer practical guidance for educators, researchers, and institutions committed to fostering more responsible, human-centered AI use in media and society.

## PART I. PREPARATION & BACKGROUND
**Where is the current (Gen)AI status in society?**
We are beyond AI's versatile use phase. Integrating AI into societal structures continues, influenced by socio-technical visions and expectations (Bareis & Katzenbach 2022). The emerging fundamental problem areas are as follows:

- Generative AI is an evolutionary leap reshaping creativity, labor, and society (Feher 2025)
- The necessary AI infrastructure is costly and not sustainable (Varoquaux et al. 2025)
- Hundreds of possible future scenarios can apply to a problem area (Hsueh et al. 2021)
- A skills shortage is emerging with the transformation of human participation (Cukier 2019)
- Deepfakes, synthetic media, and AI-generated content undermine trust in news, politics, and emerging technologies (European Digital Media Observatory 2025)
- When GenAI co-creates ideas, decisions, or creative works, it blurs authorship, agency, and responsibility (Feher et al. 2024)
- Constant ethical dilemmas about "should we use this AI?" overwhelm citizens, educators, and institutions (Brennan 2025)

**AI control in media and info-communication**
We are beyond the phase of celebrating AI's versatility. Generative AI represents an evolutionary leap reshaping creativity, labor, and society (Feher 2025), yet it demands costly, unsustainable infrastructure (Varoquaux et al. 2025) and introduces a wide range of possible future scenarios (Hsueh et al. 2021). Human roles are shifting, creating a growing skills shortage (Cukier 2019), while deepfakes and synthetic media erode trust in politics and news (EDMO 2025). Co-created content blurs authorship and responsibility (Feher et al. 2024), and constant ethical dilemmas overwhelm educators and institutions (Brennan 2025). As socio-technical imaginaries drive this transformation (Bareis & Katzenbach 2022), tackling disparities and ethical risks is essential (Koves et al. 2024).

**Generative AI, synthetic worlds and education challenges**
At the same time, generative AI is emerging as an evolutionary leap that reshapes creativity, labor, and society (Feher 2025), while requiring costly and potentially unsustainable infrastructure (Varoquaux et al. 2025). The uncertainty surrounding its societal impact is amplified by the vast number of plausible future scenarios it generates (Hsueh et al. 2021). Human participation is undergoing transformation, contributing to a growing skills shortage (Cukier 2019). Trust in media and political communication is increasingly undermined by deepfakes and synthetic content (European Digital Media Observatory 2025), while AI co-creation blurs authorship, agency, and responsibility (Feher et al. 2024). As a result, educators and institutions face constant ethical dilemmas that challenge decision-making and governance (Brennan 2025).
The widespread availability of affordable AI services is already reshaping media and communication through applications such as robot journalism, automated news delivery, and synthetic content, with deepfakes posing societal challenges (Winiarska-Brodowska & Feher 2025; Feher & Katona 2021). As these systems become interconnected—merging generative AI with immersive environments like metaverses—their societal impact intensifies. However, key breakthroughs remain contingent on addressing data protection, transparency, and persistent ethical concerns (He et al. 2023; Borsci et al. 2022).

**AI risks and ethical challenges in media and society**
AI poses significant risks, including data protection, security, biases, and potential abuse (Binns 2018). Mitigating these risks requires accurate training data selection, specialized model experimentation, and human oversight. Predictable AI algorithms are essential for supporting core social functions. Addressing these issues demands critical approaches from research centers like Oxford's Ethics in AI and NYU's Center for Responsible AI, focusing on normative ethics (Sethi et al. 2022). In media and info-communication, additional risks and ethical concerns are emerging, highlighting the long-term and deepening impact of AI proliferation, as follows:

- Generative AI reshapes labor and resource dynamics. How can societies ensure equitable readiness and access? (Varoquaux et al. 2025, Cukier 2019)
- Deepfakes, synthetic media, and AI-generated content undermine trust in news, politics, and emerging technologies. How do we restore and preserve public trust? (European Digital Media Observatory 2025, Stokel-Walker 2023)
- AI-generated content blurs authorship. How do we ensure transparency and attribution? (Feher et al. 2024)
- Constant ethical dilemmas about "should we use this AI?" overwhelm citizens, educators, and institutions. How can ethical literacy and decision capacity be strengthened? (Brennan 2025)
- Personalization, recommendation, and fact-checking involve errors. How do we ensure accuracy and mitigate the loss of valuable information? (Leiser 2022)
- Synthetic and virtual worlds can lead to addiction. How to use training for social good while avoiding abuses? (Eun et al. 2023)
- Who governs AI-driven data use, privacy, and surveillance boundaries? (Jain 2025)
- AI outputs are max. 50–70% accurate. How do we reduce hallucinations and translate "social good" into coding? (Lin et al. 2021)
- Large-scale persuasion with AI. How do we regulate abuse and formulate ethical guidelines? (Goldstein et al. 2023)
- AI and intellectual property. Who is the creator, and how does it affect the cultural-creative industry? (Floridi 2023)
- Credibility and trust in AI services. What makes content credible, and how can deception be filtered? (Feher et al. 2024; Glikson & Asscher 2023)

At the workshop, we focused on the three most critical issues for debate, with additional issues supporting the arguments. The goal was to summarize and critically approach recent AI trends, their risks, and ethical concerns.

**PART II. SUMMARY OF THE SEMINAR DISCUSSION**
The seminar drew great interest, with participants primarily being doctoral students from diverse fields such as cybersecurity, military AI applications, innovation, legal regulation, national security, diplomacy, and IT.
The seminar began with a summary of professional material, followed by topic selection based on participant interests in risks and ethical issues. Eleven core questions were proposed, with one additional topic briefly mentioned. Of the eleven issues, six were relevant to multiple participants, and one was widely discussed by almost all.

Key topics included the spread of fake and synthetic content, reputational and legal risks, mixed realities and personal data use, authorship and trust, large-scale influence campaigns, and technology for social good. Participants concluded that data drives business models, highlighting primary risks such as security challenges, abuse potential, information overload, and opaque data use even in paid services.

Regarding fact-checking, participants acknowledged its significance but noted its limited application in daily media consumption, pointing to the erosion of public trust and the need for stronger information resilience. Economic inequality, power asymmetries, and the ongoing tension between human self-worth and machine efficiency were also discussed.

The importance of critical and ethical thinking was consistently emphasized, with one participant citing Asimov in relation to the evolving co-adaptation between humans and machines. A widely relevant issue was the drastic expansion of the AI user base and the growing concern about whether users can be effectively and ethically trained to navigate emerging technologies. The challenges of authorship, blurred responsibility, and ethical decision fatigue were also addressed.

*A fundamental conclusion was that while current paradigms still center around security awareness, a significant leap is already underway. One participant emphasized the role of scientific elites and science communication in supporting independent education beyond commercial interests. Though representing science as a Ph.D. student, the participant did not overstate their role. Building on this point, others noted that the rapid spread of generative AI and its varied, scenario-based uses also challenge academic integrity, scientific equity, and the accessibility of high-quality analytical tools across regions.*

*The issue of structural inequality also emerged. Questions were raised about the availability of software and AI-based analysis in disadvantaged regions and about how input-output relationships are managed by AI service providers, including in commercial models.*

*In the ethics domain, participants reflected on the constancy of human nature alongside shifting societal functions. The pressures of continuous ethical decision-making were described as exhausting for institutions and individuals alike.*

*At the end of the debate and based on the above, participants identified five cornerstones for reducing ethical risks:*

- *Public education and early socialization into technology play a primary role starting from kindergarten.*
- *Business and service environments should be actively involved in education and receive training themselves.*
- *Universities are advised to teach these topics to maintain their credibility and social function.*
- *The role of parents and schools is critical in fostering ethical behavior and responsible technology use.*
- *The older generation needs more thorough preparation to recognize and avoid abuses.*
- *Participants emphasized the importance of continuous, society-wide educational sensitization.*
- *The majority expressed a moderately technophilic stance, fundamentally combined with a critical, ethically conscious perspective.*

Consequently, the consensus highlights the vital importance of a comprehensive technology education strategy beginning in early childhood. This must be critical, ethical, and inclusive—engaging all layers of society: businesses, schools, universities, families, and older generations alike. Summarizing the suggested actions, a multi-level lifelong learning approach is needed to foster ethical and critical engagement with technology, ensuring adaptability in the face of AI's rapid evolution.

**CONSCLUSION**
Generative AI marks an evolutionary leap, transforming media, authorship, and public trust. As synthetic content reshapes how we consume, create, and validate knowledge, societal readiness is no longer optional—it is urgent. The experimental PhD seminar revealed critical gaps in ethical capacity, technological literacy, and institutional adaptability.

The contribution lies in foregrounding actionable insights: rethinking education across all generations, integrating ethical AI training into business and public institutions, and critically examining power asymmetries in access, data use, and authorship. These results emerged specifically in the context of society and media-communication, where generative AI's rapid proliferation is most visible and emotionally impactful. Media content shapes public narratives, while communication platforms amplify both risks and hopes. This visibility accelerates ethical questioning and trust crises, making society and media the frontline in understanding and addressing the deeper consequences of synthetic knowledge production.

This framework is immediately usable for educators designing curricula, policymakers shaping AI governance, researchers exploring socio-technical systems, and tech developers aiming for responsible innovation. Its strength is in connecting ethical foresight with structural recommendations: start early, train widely, govern transparently. These findings offer not only reflection but direction—toward building resilient societies capable of navigating the risks and possibilities of AI-driven futures.

**REFERENCES**

Bareis, J., & Katzenbach, C. (2022). Talking AI into being: The narratives and imaginaries of national AI strategies and their performative politics. Science, Technology, & Human Values, 47(5), 855-881. https://doi.org/10.1177/01622439211030007

Binns, R. (2018, January). Fairness in machine learning: Lessons from political philosophy. In Conference on fairness, accountability and transparency (pp. 149-159). PMLR.

Borsci, S., Lehtola, V. V., Nex, F., Yang, M. Y., Augustijn, E. W., Bagheriye, L., ... & Zurita-Milla, R. (2022). Embedding artificial intelligence in society: looking beyond the EU AI master plan using the culture cycle. AI & Society, 1-20. https://doi.org/10.1007/s00146-021-01383-x

Brennan, K., Kak, A. & Myers West, S (2025) Artificial Power: AI Now 2025 Landscape. AI Now Institute, June 3, 2025, https://ainowinstitute.org/2025-landscape.

Cukier, W. (2019). Disruptive processes and skills mismatches in the new economy: Theorizing social inclusion and innovation as solutions. Journal of Global Responsibility. Vol. 10 No. 3, pp. 211-225. https://10.1108/JGR-11-2018-0079

Eun, S. J., Kim, E. J., & Kim, J. (2023). Artificial intelligence-based personalized serious game for enhancing the physical and cognitive abilities of the elderly. Future Generation Computer Systems, 141, 713-722. https://doi.org/10.1016/j.future.2022.12.017

European Digital Media Observatory (2025) Implementing the EU Code of Practice on Disinformation. https://edmo.eu/wp-content/uploads/2025/06/EDMO-Report-–-Implementing-the-EU-Code-of-Practice-on-Disinformation.pdf?utm_source=chatgpt.com

Feher, K. (2025) Generative AI, Media, and Society. New York, London: Routledge.

Feher, K. (2024). Exploring AI media. Definitions, conceptual model, research agenda. Journal of Media Business Studies, 1–24. https://doi.org/10.1080/16522354.2024.2340419

Feher, K., Vicsek, L., Deuze, M. (2024) Modeling AI Trust for 2050, AI and Society, Early Cite

Feher, K., & Katona, A. I. (2021). Fifteen Shadows of socio-cultural AI: A Systematic Review and future perspectives. Futures, 132, 102817. https://doi.org/10.1016/j.futures.2021.102817

Feher, K., & Veres, Z. (2022). Trends, risks and potential cooperations in the AI development market: expectations of the Hungarian investors and developers in an international context. International Journal of Sociology and Social Policy. Vol. 43 No. 1/2, pp. 107-125. https://doi.org/10.1108/IJSSP-08-2021-0205

Floridi, L. (2023, February 16). AI as Agency without Intelligence: On ChatGPT, large language models, and other generative models. Philosophy and Technology. https://doi.org/10.1007/s13347-023-00621-y

Gering, Z., Feher, K., Harmat, V., & Tamassy, R. (2025). Strategic organisational responses to generative AI-driven digital transformation in leading higher education institutions. International Journal of Organizational Analysis, 33(12), 132-152. https://doi.org/10.1108/IJOA-09-2024-4850

Goldstein, J. A., Sastry, G., Musser, M., DiResta, R., Gentzel, M., Sedova, K. (2023) Forecasting Potential Misuses of Language Models for Disinformation Campaigns—and How to Reduce Riska. https://arxiv.org/pdf/2301.04246.pdf

Glikson, E., & Asscher, O. (2023). AI-mediated apology in a multilingual work context: Implications for perceived authenticity and willingness to forgive. Computers in Human Behavior, 140, 107592.

Guzman, A. L., & Lewis, S. C. (2020). Artificial intelligence and communication: A Human–Machine Communication research agenda. New Media & Society, 22(1), 70-86. https://doi.org/10.1177/1461444819858691

He, L., Liu, K., He, Z., & Cao, L. (2023). Three-dimensional holographic communication system for the metaverse. Optics Communications, 526, 128894. https://doi.org/10.1016/j.chb.2022.107592

Hsueh, S. L., Zhou, B., Chen, Y. L., & Yan, M. R. (2021). Supporting technology-enabled design education and practices by DFuzzy decision model: applications of cultural and creative product design. International Journal of Technology and Design Education, 1-18. https://doi.org/10.1007/s10798-021-09681-7

Jackson, D., & Latham, A. (2022). Talk to The Ghost: The Storybox methodology for faster development of storytelling chatbots. Expert Systems with Applications, 190, 116223. https://doi.org/10.1016/j.eswa.2021.116223

Hepp, A. (2020). Deep mediatization: Key ideas in media & cultural studies. London: Routledge.

Jain, A. (2025) AI and Privacy 2024 to 2025. https://cloudsecurityalliance.org/blog/2025/04/22/ai-and-privacy-2024-to-2025-embracing-the-future-of-global-legal-developments?utm_source=chatgpt.com#

Kalpokas, I., & Kalpokiene, J. (2021). Synthetic media and information warfare: Assessing potential threats. The Russian Federation in Global Knowledge Warfare: Influence Operations in Europe and Its Neighbourhood, 33-50.

Koves, A., Feher, K., Vicsek, L., & Fischer, M. (2024). Entangled AI: artificial intelligence that serves the future. AI & Society, 1-12. https://doi.org/10.1007/s00146-024-02037-4

Kovtun, V., Izonin, I., & Gregus, M. (2022). The model of functioning of the centralized wireless information ecosystem is focused on multimedia streaming. Egyptian Informatics Journal, 23(4), 89-96. https://doi.org/10.1016/j.eij.2022.06.009

Leiser, M. R. (2022). Bias, journalistic endeavors, and the risks of artificial intelligence. In Artificial Intelligence and the Media (pp. 8-32). Edward Elgar Publishing. HTTPS://10.4337/9781839109973.00007

Lin, S., Hilton, J., & Evans, O. (2021). Truthfulqa: Measuring how models mimic human falsehoods. https://arxiv2109.07958

Matthias, A. (2020). Dignity and Dissent in Humans and Non-humans. Science and Engineering Ethics. 26, 2497-2510. https://doi.org/10.1007/s11948-020-00245-x

Newlands, G. (2021). Lifting the curtain: Strategic visibility of human labour in AI-as-a-Service. Big Data & Society, 8(1), 20539517211016026. https://doi.org/10.1177/2053951721101

Muller, M., Chilton, L. B., Kantosalo, A., Martin, C. P., & Walsh, G. (2022, April). GenAICHI: Generative AI and HCI. In CHI Conference on Human Factors in Computing Systems Extended Abstracts (pp. 1-7).

Sethi, A., Tangri, T., Puri, D., Singh, A., & Agrawal, K. (2022). Knowledge management and ethical vulnerability in AI. AI and Ethics, 1-8. https://doi.org/10.1007/s43681-022-00164-6

Stokel-Walker, C. (2023) ChatGPT listed as author on research papers: many scientists disapprove. Nature. PMID: 36653617. https://doi.org/10.1038/d41586-023-00107-z

Van Dijck, J. (2021). Seeing the forest for the trees: Visualizing platformization and its governance. New Media & Society, 23(9), 2801-2819. https://doi.org/10.1177/1461444820940293

Varoquaux, G., Luccioni, S., & Whittaker, M. (2025, June). Hype, Sustainability, and the Price of the Bigger-is-Better Paradigm in AI. In *Proceedings of the 2025 ACM Conference on Fairness, Accountability, and Transparency* (pp. 61-75). https://arxiv.org/abs/2409.14160v2

Winiarska-Brodowska, M., & Feher, K. (2025). Socio-cultural artificial intelligence (SCAI) and journalism: a transdisciplinary perspective in media and communication studies. In Algorithms, Artificial Intelligence and Beyond (pp. 105-120). Routledge.

*Data of the exploratory seminar*
- Seminar leader: Dr. habil Katalin Feher, Associate Professor, Ludovika UPS
- Participants: PhD students of the Doctoral School of Public Administration Sciences at Ludovika University of Public Service
- Acknowledgment: UNKP Bolyai+, Grant number: UNKP-22-5-NKE-87